\documentclass[conference, 10pt]{IEEEtran}

\usepackage{cite}
\usepackage{url}
\usepackage{graphicx}
\usepackage{color}
\usepackage{placeins}
\usepackage{float}
\usepackage{tabularx,colortbl}
\usepackage{ifthen}

\makeatletter

\newcounter{author}
\renewcommand{\author}[2][]{
   \stepcounter{author}
   \@namedef{author@\theauthor}{#2}
   \@namedef{authorlabel@\theauthor}{#1}
}

\newcounter{address}
\newcommand{\address}[2][]{
   \stepcounter{address}
   \@namedef{address@\theaddress}{#2}
   \@namedef{addresslabel@\theaddress}{#1}
}

\newcommand{\alsep}{and}

\def\newmaketitle{\par%
  \begingroup%
  \normalfont%
  \def\thefootnote{}
  \def\footnotemark{}
  \let\@makefnmark\relax
  \footnotesize
  \footnotesep 0.7\baselineskip
  \normalsize%
  \twocolumn[\thenewmaketitle\@IEEEaftertitletext]%
  \if@IEEEusingpubid
     \enlargethispage{-\@IEEEpubidpullup}%
  \fi
  \endgroup
  \setcounter{footnote}{0}\let\maketitle\relax\let\@maketitle\relax
  \gdef\@thanks{}%
  \let\thanks\relax}

\def\thenewmaketitle{
  \newpage
  \begin{center}%
    \vskip0.2em{\Huge\@IEEEcompsoconly{\sffamily}\@IEEEcompsocconfonly{\normalfont\normalsize\vskip 2\@IEEEnormalsizeunitybaselineskip
   \bfseries\large}\@title\par}\vskip1.0em\par%
    \vspace{1ex}
    \newcounter{c@author}
    \newcounter{c@tmp}
    \ifthenelse{\value{author}=2}{%
      \newcommand{\liand}{ and }}{%
      \newcommand{\liand}{, and }}
    \ifthenelse{\value{address}<2}{%
      \@nameuse{author@1}%
      \stepcounter{c@author}%
      \whiledo{\value{c@author}<\value{author}}{%
        \setcounter{c@tmp}{\value{author}}%
        \addtocounter{c@tmp}{-\value{c@author}}%
        \ifthenelse{\value{c@tmp}=1}{%
          \renewcommand{\alsep}{\liand}}{\renewcommand{\alsep}{, }}%
        \stepcounter{c@author}\alsep \@nameuse{author@\thec@author}}\\%
    }
    {
      \@nameuse{author@1}${}^{(\ref{\@nameuse{authorlabel@1}})}$%
      \stepcounter{c@author}%
      \whiledo{\value{c@author}<\value{author}}{%
      \setcounter{c@tmp}{\value{author}}%
      \addtocounter{c@tmp}{-\value{c@author}}%
      \ifthenelse{\value{c@tmp}=1}{%
        \renewcommand{\alsep}{\liand}}{\renewcommand{\alsep}{, }}%
      \stepcounter{c@author}\alsep \@nameuse{author@\thec@author}%
        ${}^{(\ref{\@nameuse{authorlabel@\thec@author}})}$%
      }
    }
    \vspace{0.2ex}

    \ifthenelse{\value{address}>0}{%
      \ifthenelse{\value{address}=1}{
        {\@nameuse{address@1}}
      }
      {
        \newcounter{c@address}

        \begin{center}
        \whiledo{\value{c@address}<\value{address}}
        {
          \refstepcounter{c@address}
            ${}^{(\thec@address)}$\,%
              \label{\@nameuse{addresslabel@\thec@address}}%
              \@nameuse{address@\thec@address}\\ %
        }
        \end{center}
      } 
    }
    {
      \relax
    }
  \end{center}
}

\makeatother

\title{An Electronically Tunable 28-34 GHz 2-D Steerable Leaky Wave Antenna}

\author[org1]{Mahdi Alesheikh}
\author[org1]{Md Hedayatullah Maktoomi}
\author[org2]{Soheil Saadat}
\author[org1]{Hamidreza Aghasi}
\address[org1]{The University of California, Irvine, CA, 92617 USA}
\address[org2]{MFLEX Inc., Irvine, CA,  92617, USA}

\begin{document}

\newmaketitle

\begin{abstract}
 In this paper, a 2-D beam steering mm-wave antenna based on the leaky wave configuration is presented. Microstrip leaky wave antennas are known to exhibit beam rotation by changing the frequency. In this work, the microstrip leaky wave antenna is adopted and co-integrated with electronically tunable board components that periodically load the antenna. By independent control of variable capacitors and diodes, single-frequency 2-D beam steering across the bandwidth is achieved. The proposed antenna is fabricated in Rogers printed circuit board technologies and the simulation results exhibit a peak realized gain of 8 dBi, radiation bandwidth of 28-34 GHz, radiation efficiency of more than {80}\%, and more than 90$^\circ$ and 70$^\circ$ of beam rotation in the $\phi$ and $\theta$ directions.
\end{abstract}

\section{Introduction}
With advancements in the next generation of communication systems and automotive radars \cite{shahriar,xuyang,mahdi}, the development of 2-D beam steering has gained a growing interest. In automotive radar systems, the blind spots in the elevation and azimuth directions should be mitigated by deploying 2-D beam steering. Conventional phased-array architectures with 2-D beam steering in large antenna arrays deploy digital beam forming at radio frequencies and analog beamforming at mm-wave frequencies. Analog beamforming techniques are prone to losses from the phase shifters and feed distribution networks and digital beamforming techniques suffer from processing delays and computational burden. Therefore, alternatives for low-loss 2-D beam steering that can be integrated with nanoscale CMOS ICs at mm-wave frequencies are needed. 


A large body of prior work on passive approaches to steer the beam of antenna have revolved around electromechanical approaches \cite{emech1} which are hard to integrate with IC-based mm-wave systems. Frequency-dependent and single-frequency steering of beams in planar structures such as travelling antennas has also been studied to enable heterogeneous integration between the CMOS IC and the antenna, \cite{diffreq,fixedfreq}. However, most of these works achieve a single-axis rotation of beam and large 2-D arrays with sophisticated feed networks are required to adopt them in phased arrays. In this paper, we adopt the half-width microstrip leaky wave antenna (HWMLWA) and co-design it with a periodic network of electronically tunable passive components, i.e., diodes and variable capacitors (varactors). By leveraging the impedance variation of these components that directly impacts the profile of travelling \textit{E}-field on the antenna, more than 70$^\circ$ of beam rotation is achieved in both directions of the HWMLWA.

The rest of this paper is organized as follows. The fabrication details and operation principle of the antenna are presented in Section \ref{sec2}. The simulation results are provided in Section \ref{sec3} and the paper is concluded in Section \ref{sec4}.
\section{Design of the Antenna}
\label{sec2}
\vspace{-0.15in}
\begin{figure}[!h]
    \centering
    \includegraphics[width=0.95\linewidth]{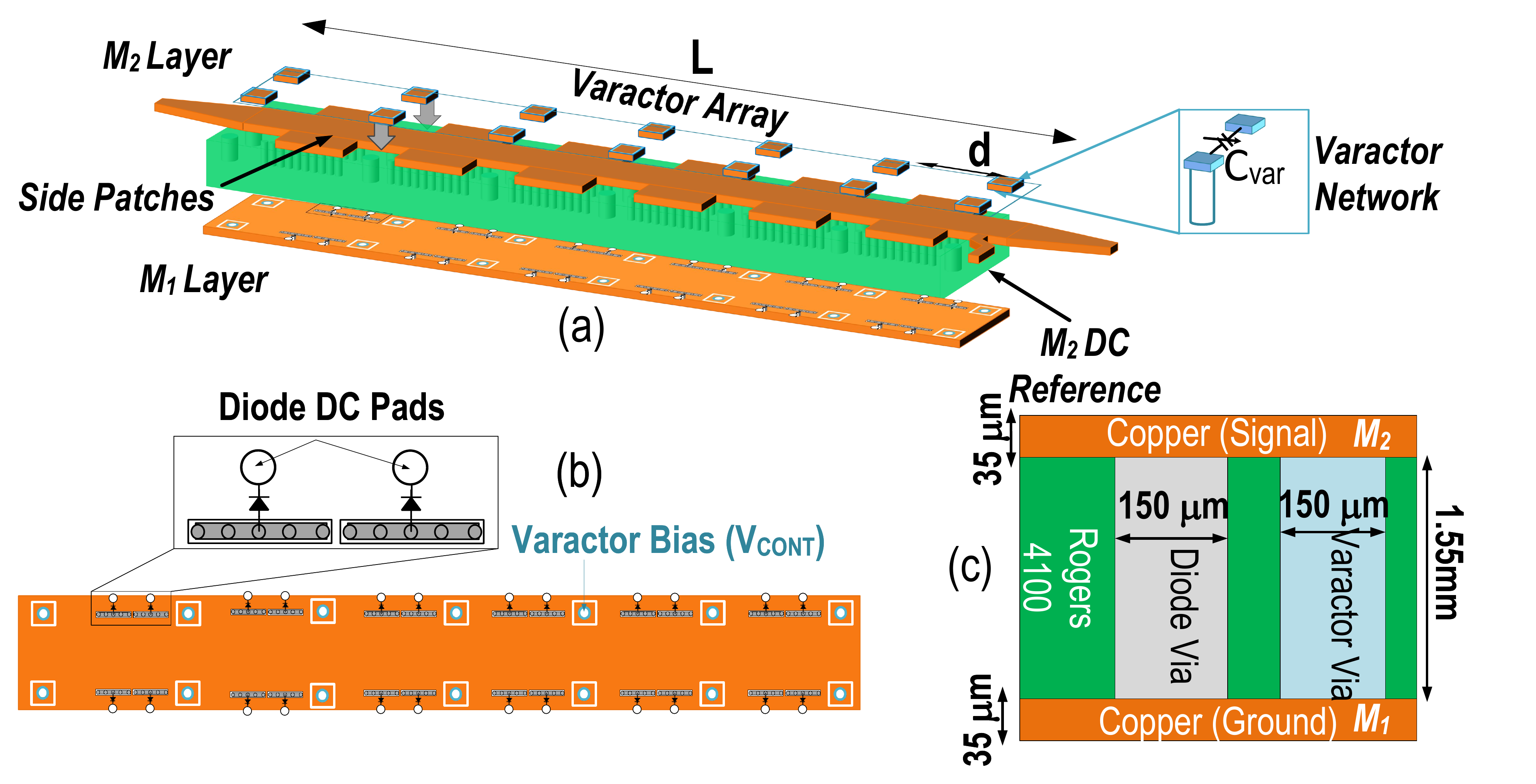}
    \vspace{-0.15in}
    \caption{(a) 3-Dimensional cross-section view of the proposed antenna, (b) backside view of $M_1$, (c) cross-section view of the design.}
    \label{fig:1}
\end{figure}
The proposed antenna structure is shown in Fig \ref{fig:1}. The main travelling wave antenna is a HWMLWA with \textit{L}=39 mm (45 mm from port 1= signal to port 2= 50 $\Omega$) and a tapered width from 2 mm (at the location of 50 $\Omega$ connectors) to 2.5 mm in the middle of the line. The antenna is fabricated on RO-4350B dielectric with
1.55 mm height and $\epsilon_r=$3.66/DF=0.0037. 

\begin{figure}
    \centering
    \includegraphics[width=1\linewidth]{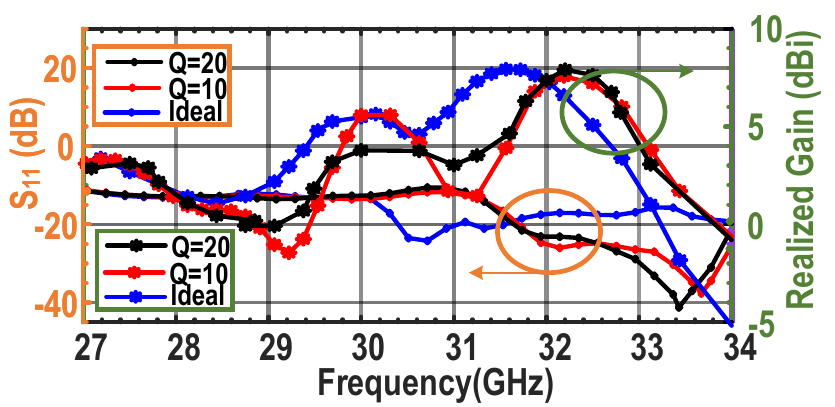}
    \vspace{-0.25in}
    \caption{Impact of \textit{Q} variation on the performance of the proposed antenna}
    \label{fig:2}
\end{figure}

In a conventional MLWA, the radiation angle, $\theta_f$, along the wave propagation axis tilts by changing frequency according to $\sin (\theta_f)=\beta\left(f\right)/K_0\left(f\right)$, where $\beta$ and $K_0$ are the phase and propagation constants, respectively. To enable single-frequency beam rotation in the $\theta$ direction, we deploy 6 periodic varactor loadings on each side of the HWMLWA with spacings of \textit{d}=5 mm. To define the DC state of top signal metal layer, a $\lambda/4$ DC line near port 2 is routed on $M_2$ layer. The other end of each varactor is routed through vias and connected to $V_{CONT}$ at the back side of lower metal ($M_1$), as shown in Fig. 1 (b). To increase the degrees of freedom, separate $V_{CONT}$ for each varactor is considered. The change of varactors by their control voltages allows to rotate the angle of beam at a single frequency, $f_0$, based on a modified equation, i.e., $\sin (\theta_{f_0,V_{CONT,1},\cdots V_{CONT,12}})=\beta\left({f_0,V_{CONT,1}, \cdots V_{CONT,12}}\right)/K_0\left(f_0\right)$. One of the important considerations to deploy varactors is the quality factor (\textit{Q}) degradation at mm-wave frequencies. For the selected varactor models (MACOM MTV4030), we evaluated the effect of \textit{Q} on the antenna impedance matching and realized gain, as demonstrated in Fig. \ref{fig:2}. It is evident that for the alternation of \textit{Q} between 10 and 20 (achievable for the selected model), the realized gain of the antenna does not change significantly. Moreover, for the target frequency range (28-34 GHz) $S_{11}$ remains well below -10 dB when the varactor \textit{Q} changes.

To enable the beam rotation in the $\phi$ (y-axis) direction one idea is to alter the grounded side in the HWMLWA. However, a solid ground creation on either side creates a binary $\phi$ rotation, e.g., -60$^\circ$ and $+$60$^\circ$, which is not practical for radar applications. To enhance the $\phi$ rotation resolution, we deploy 6 quasi-patch periodic loadings with 5 mm spacing on each side that are controlled by low-loss diode networks (MACOM MMSPN050 series). Each patch is vertically connected to an anode reference plane (isolated from the ground) by parallel combination of 5 vias with 150 $\mu$m spacing to mitigate the inductive and resistive effect of the vias. By connecting the anode of each diode to a reference plane and the cathode to an external DC voltage, each individual patch can be short-circuited or open-circuited with respect to ground. By controlling the number of patches on each side that are short-circuited, the intensity of electric field on each side is varied hence translating to a $\phi$ rotation of the beam.

\begin{figure}
    \centering
    \includegraphics[width=0.9\linewidth]{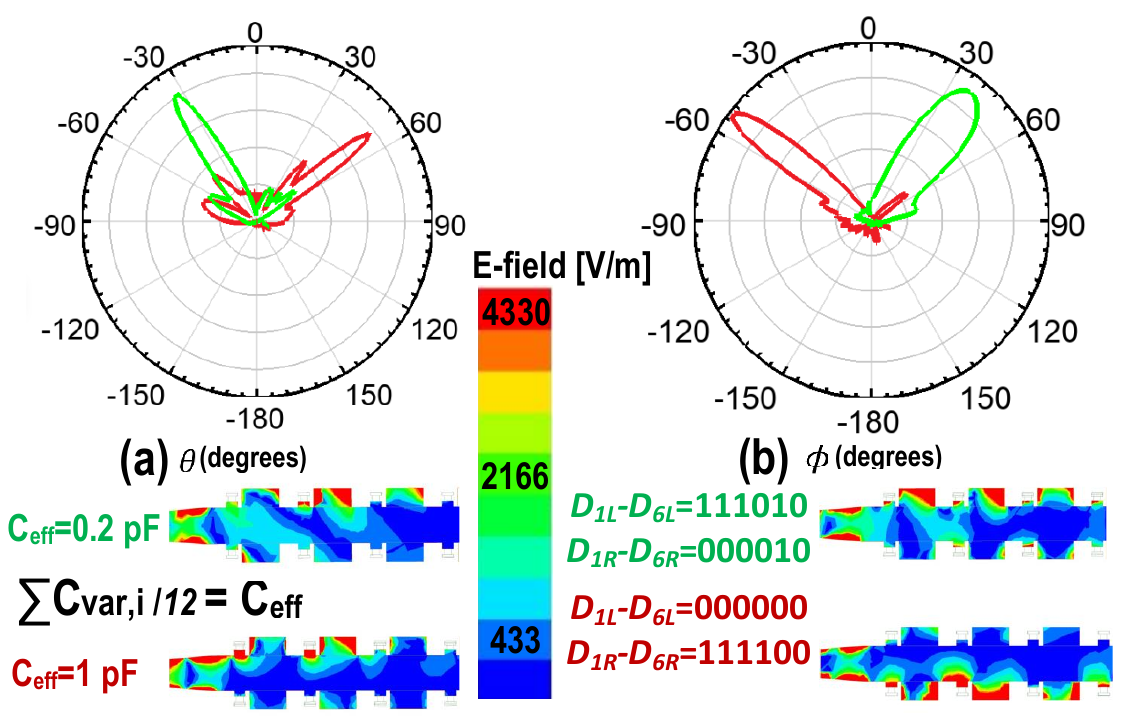}
    \vspace{-0.15in}
    \caption{Steering the beam in (a) $\theta$ and (b) $\phi$ directions as well as corresponding \textit{$\vec{E}$} profiles by altering the varactors and diode states at 31 GHz.}
    \label{fig:3}
\end{figure}


   \begin{figure}
       \centering
       \includegraphics[width=0.9\linewidth]{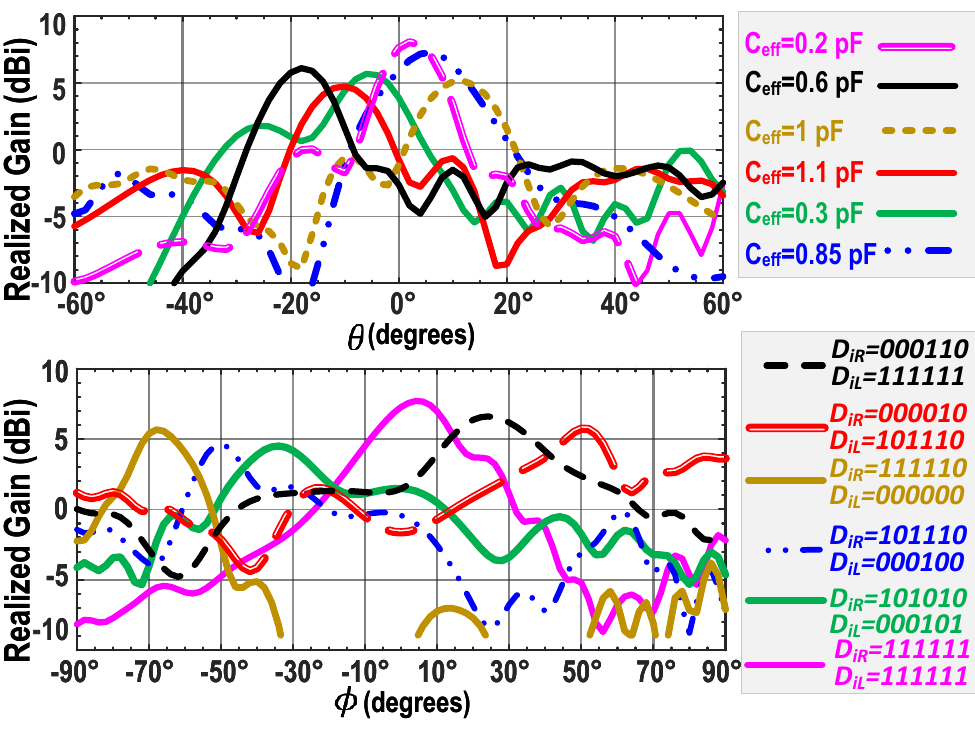}
       \vspace{-0.15in}
       \caption{Rotation of beam at 32 GHz in the (up) $\theta$ axis and (bottom) $\phi$ axis.}
       \label{fig:enter-label}
   \end{figure}
\section{Simulation results}
\label{sec3}
Fig. \ref{fig:3} shows the rotation of beam at 31 GHz in both $\phi$ and $\theta$ directions. The antenna achieves $\theta$ rotation from -34$^\circ$ to +52$^\circ$ by changing $C_{eff}$ between 0.2 pF and 1 pF and $\phi$ rotations from -52$^\circ$ to 38$^\circ$ by controlling the state of patches on the left side ($D_{1L}-D_{6L}$) and right side ($D_{1R}-D_{6R}$), respectively. The \textit{E}-field profiles corresponding to each scenario are also depicted in Fig. \ref{fig:3}. For another frequency (32 GHz) in Fig. \ref{fig:enter-label}, finer steps of $\phi$ and $\theta$ tilting and their corresponding $C_{eff}$ and $D_{iL}/D_{iR}$ states are illustrated. The results show a beam steering precision close to 10$^\circ$ in both directions with gain fluctuations less than 2 dB. The measurements of the fabricated antenna is ongoing and authors plan to provide measurement results in early future.   


\section{Conclusion}
\label{sec4}
Millimeter-wave antennas with 2-D beam steering, low profile, and low-cost fabrication are required for the next generation of communication and radar systems. In this paper, a 2-D beam steering half-width leaky wave antenna is presented to control the beam in azimuth and elevation directions. This electronically controlled antenna for mm-wave radar systems at the 28-34 GHz band achieves a peak realized gain of 8 dBi and more than 70$^\circ$ of rotation in both $\theta$ and $\phi$ axis. The co-integration of this antenna with mm-wave radar ICs helps to remove lossy phase-shifters from the front-ends without degrading the range and precision of beam steering.

\bibliographystyle{IEEEtran}
\bibliography{reference}

\begin{thebibliography}{1}
\providecommand{\url}[1]{#1}
\csname url@samestyle\endcsname
\providecommand{\newblock}{\relax}
\providecommand{\bibinfo}[2]{#2}
\providecommand{\BIBentrySTDinterwordspacing}{\spaceskip=0pt\relax}
\providecommand{\BIBentryALTinterwordstretchfactor}{4}
\providecommand{\BIBentryALTinterwordspacing}{\spaceskip=\fontdimen2\font plus
\BIBentryALTinterwordstretchfactor\fontdimen3\font minus \fontdimen4\font\relax}
\providecommand{\BIBforeignlanguage}[2]{{%
\expandafter\ifx\csname l@#1\endcsname\relax
\typeout{** WARNING: IEEEtran.bst: No hyphenation pattern has been}%
\typeout{** loaded for the language `#1'. Using the pattern for}%
\typeout{** the default language instead.}%
\else
\language=\csname l@#1\endcsname
\fi
#2}}
\providecommand{\BIBdecl}{\relax}
\BIBdecl

\bibitem{shahriar}
M.~Yaghoobi, M.~H. Kashani, M.~Yavari, and S.~Mirabbasi, ``A 56-to-66 ghz cmos low-power phased-array receiver front-end with hybrid phase shifting scheme,'' \emph{IEEE Transactions on Circuits and Systems I: Regular Papers}, vol.~67, no.~11, pp. 4002--4014, 2020.

\bibitem{xuyang}
X.~Liu, M.~H. Maktoomi, M.~Alesheikh, P.~Heydari, and H.~Aghasi, ``A 49-63 ghz phase-locked fmcw radar transceiver for high resolution applications,'' in \emph{ESSCIRC 2023-IEEE 49th European Solid State Circuits Conference (ESSCIRC)}.\hskip 1em plus 0.5em minus 0.4em\relax IEEE, 2023, pp. 509--512.

\bibitem{mahdi}
M.~Alesheikh, R.~Feghhi, F.~M. Sabzevari, A.~Karimov, M.~Hossain, and K.~Rambabu, ``Design of a high-power gaussian pulse transmitter for sensing and imaging of buried objects,'' \emph{IEEE Sensors Journal}, vol.~22, no.~1, pp. 279--287, 2022.

\bibitem{emech1}
A.~A. Baba, R.~M. Hashmi, K.~P. Esselle, M.~Attygalle, and D.~Borg, ``A millimeter-wave antenna system for wideband 2-d beam steering,'' \emph{IEEE Transactions on Antennas and Propagation}, vol.~68, no.~5, pp. 3453--3464, 2020.

\bibitem{diffreq}
H.~Saeidi, S.~Venkatesh, X.~Lu, and K.~Sengupta, ``Thz prism: One-shot simultaneous localization of multiple wireless nodes with leaky-wave thz antennas and transceivers in cmos,'' \emph{IEEE Journal of Solid-State Circuits}, vol.~56, no.~12, pp. 3840--3854, 2021.

\bibitem{fixedfreq}
D.~K. Karmokar, D.~N.~P. Thalakotuna, K.~P. Esselle, L.~Matekovits, and M.~Heimlich, ``Reconfigurable half-width microstrip leaky-wave antenna for fixed-frequency beam scanning,'' in \emph{2013 7th European Conference on Antennas and Propagation (EuCAP)}, 2013, pp. 1314--1317.

\end{thebibliography}

\end{document}